\documentclass{PoS}

\usepackage{amsmath,amstext,amsfonts,amsbsy,amssymb,amscd,bbm,epsfig}

\title{Anomalous dimensions of four-fermion operators from conformal EWSB dynamics\footnotetext{Preprint n.: IFT-UAM/CSIC-13-122, FTUAM-13-35}}

\ShortTitle{Anomalous dimensions of four-fermion operators from conformal EWSB dynamics}

\author{Luigi Del Debbio\\
        The Higgs Centre for Theoretical Physics\\ The University of Edinburgh, Edinburgh, UK\\
        E-mail: \email{Luigi.Del.Debbio@ed.ac.uk}}

\author{Liam Keegan\\
        Instituto de F\'{\i}sica Te\'orica UAM/CSIC\\ Universidad Aut\'onoma de Madrid, Cantoblanco 28049 Madrid, Spain\thanks{Address after November 1, 2013: CERN, Physics Department, 1211 Geneva 23, Switzerland}\\
        E-mail: \email{liam.keegan@cern.ch}}

\author{\speaker{Carlos Pena}\\
        Departamento de F\'{\i}sica Te\'orica and Instituto de F\'{\i}sica Te\'orica UAM/CSIC\\ Universidad Aut\'onoma de Madrid, Cantoblanco 28049 Madrid, Spain \\
        E-mail: \email{carlos.pena@uam.es}}

\abstract{(Quasi)conformal scaling of composite operators from a strongly coupled EWSB
dynamics helps to produce the characteristic hierarchies exhibited by the flavour couplings
of the SM. It is however crucial to ensure that specific models satisfy bounds on Higgs and
flavour dynamics; this in turn requires to control not only the anomalous dimensions of bilinears,
but also those of higher-dimensional operators. We report on an ongoing effort to determine
four-fermion operator anomalous dimensions, via Schr\"odinger Functional techniques,
in the benchmark scenario of Minimal Walking Technicolour.}

\FullConference{31st International Symposium on Lattice Field Theory - LATTICE 2013\\
		July 29 - August 3, 2013\\
		Mainz, Germany}

\begin{document}

\vspace{-1mm}\section{Motivation}

\noindent
One of the most interesting alternatives to Electroweak Symmetry Breaking (EWSB)
by a weakly-coupled fundamental scalar field is to hypothetise the appearance
of strong dynamics at or above the electroweak scale~\cite{TC}.
In order to be compatible with the tight constraints imposed by the flavour
structure of the Standard Model, one particularly appealing possibility is
that the symmetry breaking sector involves an (approximate) conformal
symmetry that allows for a natural accommodation of flavour hierarchies
in the dynamics, with the added possibility of having a light dilaton
playing the role of the Higgs boson~\cite{WTC}.
In recent years, various candidate field theories
to display such a behaviour have been studied extensively on the lattice,
using the sophisticated tools developed for QCD (see e.g.~\cite{kuti}
for review at this conference).

One of the most important building blocks for model building in this context is the
determination of the anomalous dimensions of composite operators.
Apart from the obvious case of the techniquark mass anomalous dimension,
it is also important to have information about four-fermion operators,
that can have a relevant role not only in the flavour sector, but also
e.g. in corrections to the Higgs couplings~\cite{Rattazzi:2008pe}. In this work
we extend well-known QCD results on renormalisation of four-fermion
operators to theories with adjoint fermion fields, and perform a
preliminary study for the case of Minimal Walking Technicolour,
building upon previous studies of this theory~\cite{MWTC1,MWTC2}.

\vspace{-1mm}\section{Renormalisation of four-fermion operators}

\vspace{-1mm}\subsection{Complete basis of operators}

\noindent
The renormalisation of four-fermion operators in vector ${\rm SU}(N)$ gauge theories with
fundamental fermion fields (e.g. QCD) is well understood. A comprehensive summary of the
main results can be found e.g. in~\cite{Donini:1999sf}. Generic Lorentz- and gauge-invariant
operators can be written as
\vspace{-1mm}\begin{gather}
\label{eq:opers}
(\bar\psi_\alpha\Gamma_1\psi_\beta)
(\bar\psi_\gamma\Gamma_2\psi_\delta)\,,~~~~~~~~~
(\bar\psi_\alpha\Gamma_1T^A\psi_\beta)
(\bar\psi_\gamma\Gamma_2T^A\psi_\delta)\,,
\end{gather}
where $\alpha,\beta,\gamma,\delta$ are flavour indices; $\Gamma_{1,2}$ are spin
matrices (possibly containing summed-over Lorentz indices);
$T^A$ are colour generators with $A=1,\ldots,N^2-1$; and parentheses
indicate traces over spin and colour.

In order to obtain a complete set of independent operators with a given
flavour structure, one first observes that there are only ten independent
spin structures $\Gamma_1 \otimes \Gamma_2$, that can be furthermore chosen
to lead to operators that transform in a well-defined way under parity:
\vspace{-1mm}\begin{gather}
\begin{split}
&\mbox{parity-even:~~}
\gamma_\mu \otimes \gamma_\mu\,,~
\gamma_\mu\gamma_5 \otimes \gamma_\mu\gamma_5\,,~
\sigma_{\mu\nu} \otimes \sigma_{\mu\nu}\,,~
\mathbf{1} \otimes \mathbf{1}\,,~
\gamma_5 \otimes \gamma_5\,; \\
&\mbox{parity-odd:~~}
\gamma_\mu \otimes \gamma_\mu\gamma_5\,,~
\gamma_\mu\gamma_5 \otimes \gamma_\mu\,,~
{\scriptstyle{{1\over 2}}}\epsilon_{\mu\nu\rho\tau}\sigma_{\mu\nu} \otimes \sigma_{\rho\tau}\,,~
\mathbf{1} \otimes \gamma_5\,,~
\gamma_5 \otimes \mathbf{1}\,,
\end{split}
\end{gather}
where $\sigma_{\mu\nu}=\frac{i}{2}[\gamma_\mu,\gamma_\nu]$.
Next, by using the colour trace identity and the spin Fierz identities
in the particle-antiparticle channel
\vspace{-1mm}\begin{gather}
\label{eq:fierz_fund}
(T^A)_{ab}(T^A)_{cd} = \frac{1}{2}\,\delta_{ad}\delta_{bc} - \frac{1}{2N}\,\delta_{ab}\delta_{cd}\,,~~~~~~~
(\Gamma_1^{(r)})_{ij}(\Gamma_2^{(r)})_{kl} = \sum_s f_{rs}(\Gamma_1^{(s)})_{il}(\Gamma_2^{(s)})_{kj}\,,
\end{gather}
where $f_{rs}$ are constants and the indices $r,s$ run over the ten spin structures above,
one can rewrite operators containing colour generators as linear combinations of operators
without colour generators (cf.~Eq.~(\ref{eq:opers})).
The conclusion is that a complete basis of ten operators that are eigenstates of Fierz
rearrangements with flavour content $(\alpha\beta\gamma\delta)$ is given by
\vspace{-1mm}\begin{align}
\label{eq:4f_1}
Q_1^\pm &\equiv {\scriptstyle{{1\over 2}}}\left[
(\bar\psi_\alpha\gamma_\mu(\mathbf{1}-\gamma_5)\psi_\beta)(\bar\psi_\gamma\gamma_\mu(\mathbf{1}-\gamma_5)\psi_\delta)
\pm (\beta \leftrightarrow \delta)
\right] \,,\\
\label{eq:4f_2}
Q_2^\pm &\equiv {\scriptstyle{{1\over 2}}}\left[
(\bar\psi_\alpha(\mathbf{1}-\gamma_5)\psi_\beta)(\bar\psi_\gamma(\mathbf{1}+\gamma_5)\psi_\delta)
\pm (\beta \leftrightarrow \delta)
\right] \,,\\
\label{eq:4f_3}
Q_3^\pm &\equiv {\scriptstyle{{1\over 2}}}\left[
(\bar\psi_\alpha\gamma_\mu(\mathbf{1}-\gamma_5)\psi_\beta)(\bar\psi_\gamma\gamma_\mu(\mathbf{1}+\gamma_5)\psi_\delta)
\pm (\beta \leftrightarrow \delta)
\right] \,,\\
\label{eq:4f_4}
Q_4^\pm &\equiv {\scriptstyle{{1\over 2}}}\left[
(\bar\psi_\alpha(\mathbf{1}-\gamma_5)\psi_\beta)(\bar\psi_\gamma(\mathbf{1}-\gamma_5)\psi_\delta)
\pm (\beta \leftrightarrow \delta)
\right] \,,\\
\label{eq:4f_5}
Q_5^\pm &\equiv {\scriptstyle{{1\over 2}}}\left[
(\bar\psi_\alpha\sigma_{\mu\nu}(\mathbf{1}-\gamma_5)\psi_\beta)(\bar\psi_\gamma\sigma_{\mu\nu}(\mathbf{1}-\gamma_5)\psi_\delta)
\pm (\beta \leftrightarrow \delta)
\right] \,,
\end{align}
where $(\beta \leftrightarrow \delta)$
refers to a flavour exchange. It is customary to use a shorthand notation
to make the spin structure and parity components explicit, viz.
$Q_1^\pm = Q_{\rm VV+AA}^\pm - Q_{\rm VA+AV}^\pm;~
Q_2^\pm = Q_{\rm VV-AA}^\pm + Q_{\rm VA-AV}^\pm;~
Q_3^\pm = Q_{\rm SS-PP}^\pm + Q_{\rm SP-PS}^\pm;~
Q_4^\pm = Q_{\rm SS+PP}^\pm - Q_{\rm SP+PS}^\pm;~
Q_5^\pm = Q_{\rm TT}^\pm + Q_{\rm T\bar T}^\pm$.

In order to carry over this result to the case of a theory with fermion fields
transforming in the adjoint representation of the ${\rm SU}(N)$ colour group,
the main difference lies in the structure of the colour trace identity and
the corresponding spin Fierz rearrangement needed, that now takes place in the
particle-particle channel,
\vspace{-1mm}\begin{gather}
\label{eq:fierz_adj}
(T^A)_{ab}(T^A)_{cd} = \delta_{ad}\delta_{bc} - \delta_{ac}\delta_{bd}\,,~~~~~~~
(\Gamma_1^{(r)})_{ij}(\Gamma_2^{(r)})_{kl} = \sum_s f'_{rs}(\Gamma_1^{(s)}C)_{ik}(C\Gamma_2^{(s)})_{lj}\,,
\end{gather}
where $C$ is the charge conjugation matrix, that satisfies
$C \gamma_\mu C^T = C^T \gamma_\mu C = -\gamma_\mu^T\,,~C^{-1}=C^T$.
In this way, operators of the form $(\bar\psi_\alpha\Gamma_1T^A\psi_\beta)(\bar\psi_\gamma\Gamma_2T^A\psi_\delta)$
can be written as linear combinations of operators of the forms
$(\bar\psi_\alpha\Gamma_1\psi_\delta)(\bar\psi_\gamma\Gamma_2\psi_\beta)$
and $(\bar\psi_\alpha\Gamma_1C\bar\psi_\gamma^T)(\psi_\delta^T C\Gamma_2\psi_\beta)$.

As will be shown below, the occurrence of this second kind of operators with adjoint
fermions does not lead to any qualitative difference with respect to the case of
fundamental fermions, since the structure of the relevant correlation functions
will be identical. Therefore, in the remainder of this section we will assume the structure in
Eqs.~(\ref{eq:4f_1}-\ref{eq:4f_5}).

\vspace{-1mm}\subsection{Operator mixing and anomalous dimensions}

\noindent
Under renormalisation, any composite operator will mix with all operators of
equal or lower engineering dimension that have the same transformation properties
under all the relevant symmetries. In the case of mixing with lower-dimensional
operators, the mixing coefficients can depend on fermion masses; this dependence
can actually be constrained by treating the fermion mass matrix as a spurion field
that transforms under the chiral group, such that the product of the operator
times mass factors has the same transformation properties as the four-fermion
operator of interest. The amount of flavour contained in the four-fermion operator
plays a relevant role.
For instance, the operator $(\bar s_{\rm L} \gamma_\mu d_{\rm L})(\bar s_{\rm L} \gamma_\mu d_{\rm L})$
that controls $\Delta S=2$ transitions in the Standard Model does not mix
with lower-dimensional operators, since no operator of dimension lower than
six can be written with two units of strangeness. On the opposite extreme, an operator
of the form $\sum_{\alpha,\beta=1}^{N_{\rm f}}(\bar\psi_\alpha\psi_\alpha)(\bar\psi_\beta\psi_\beta)$,
fully traced over flavour, will in general mix with several lower-dimensional operators.

The mixing structure is further complicated
by the fact that regulators usually break some of the symmetries of the target
renormalised theory; in particular, lattice regularisations break Poincar\'e
symmetry and, unless Ginsparg-Wilson quarks are used, at least part of the full
${\rm SU}(N_{\rm f})_{\rm L} \times {\rm SU}(N_{\rm f})_{\rm R}$ chiral group. We will
employ Wilson fermions, that break all axial chiral symmetries explicitly. This has two possible consequences: mixing with additional operators and less constrained mixing 
with lower-dimensional operators.

Mixing with lower-dimensional operators often leads to particularly complicated
renormalisation problems. When the combined dimensions of the operator
and the mass factors in the mixing coefficients add up to $d<6$ the missing powers of
energy will lead to a power divergence $a^{d-6}$ in the bare four-fermion
operator, that is also contained in the mixing coefficient such that the
subtracted operator diverges only logarithmically.
It can be shown that power-divergent subtractions do not contribute to anomalous
dimensions~\cite{Testa:1998ez}; as a matter of fact, if a mass-independent
scheme is used renormalisation conditions can be imposed in the chiral limit,
where subtractions that involve mass factors exactly vanish. When
an operator requires subtractions that survive the chiral limit
(as e.g. for the fully flavour-scalar operators mentioned above),
it is possible to connect it via non-singlet chiral transformations
(therefore keeping within the same chiral multiplet)
to operators with enough flavour content so as to prevent mixing below $d=6$.
It is thus convenient to extract anomalous dimensions from
operators that do not require lower-dimensional subtractions.

A convenient way to achieve this is to consider operators that are made up
by four formally distinct flavours, by setting $(\alpha\beta\gamma\delta)=(1234)$
in Eqs.~(\ref{eq:4f_1}-\ref{eq:4f_5}). A full analysis of the resulting renormalisation
problem is provided in~\cite{Donini:1999sf}. First one can use the 
${\rm SU}(4)_{\rm L} \times {\rm SU}(4)_{\rm R}$ chiral group spanned by the four
flavours to determine operator mixing in the presence of full chiral symmetry;
the outcome is that $+$ and $-$ operators
do not mix, and that each of the two resulting five-operator sets displays the renormalisation
pattern
\vspace{-1mm}\begin{gather}
\label{eq:chiral_mixing}
\left(
\begin{array}{c}
\hat Q_1^\pm \\\hat Q_2^\pm \\\hat Q_3^\pm \\\hat Q_4^\pm \\\hat Q_5^\pm
\end{array}
\right)
=
\left(
\begin{array}{ccccc}
Z_{11}^\pm & 0 & 0 & 0 & 0 \\
0 & Z_{22}^\pm & Z_{23}^\pm & 0 & 0 \\
0 & Z_{32}^\pm & Z_{33}^\pm & 0 & 0 \\
0 & 0 & 0 & Z_{44}^\pm & Z_{45}^\pm \\
0 & 0 & 0 & Z_{54}^\pm & Z_{55}^\pm
\end{array}
\right)
\left(
\begin{array}{c}
Q_1^\pm \\Q_2^\pm \\Q_3^\pm \\Q_4^\pm \\Q_5^\pm
\end{array}
\right)\,,
\end{gather}
where $\hat Q_i^\pm$ denotes a renormalised operator. Note that, furthermore,
each operator can be split into even and odd pieces under parity; 
operators with different parity do not mix, and they renormalise with the same
matrix of renormalisation constants. With Wilson fermions, the breaking of
chiral symmetry leads to all operators mixing; using an obvious vector
notation, this can be written as
\vspace{-1mm}\begin{gather}
\label{eq:wilson_mixing}
\hat{\vec Q}^\pm = Z^\pm(\mathbf{1}+\Delta^\pm)\vec Q^\pm\,, 
\end{gather}
where $\Delta^\pm$ are matrices that have non-zero entries
where $Z^\pm$ has zero entries, and vice versa.
Its elements $\Delta^\pm_{ij}$ do {\em not} depend
on the renormalisation scale, but only on the bare lattice coupling.
The structure of $\Delta^\pm$ can be further constrained by
exploiting discrete symmetries, including the discrete subgroup of flavour
exchange symmetries that are left intact by a Wilson term. One
crucial result is that parity-odd operators are protected by the so-called
$CPS$ symmetries, that combine charge conjugation, parity, and various
flavour switchings. It then follows
that, if Eq.~(\ref{eq:wilson_mixing}) is projected into well-defined
parity sectors, then $\Delta_{ij}^\pm=0$ for odd parity --- i.e. the
set of operators $Q_{\rm VA+AV}^\pm,\ldots,Q_{\rm T\tilde T}^\pm$ renormalises
as in Eq.~(\ref{eq:chiral_mixing}) even with Wilson fermions.
Since the continuum anomalous dimensions are the same for the parity-even
and parity-odd parts of the basis, the final conclusion is that the cleanest
way to determine them is to impose renormalisation conditions on parity-odd,
fully-flavoured operators in the chiral limit.

\vspace{-1mm}\section{Non-perturbative computation of anomalous dimensions}

\noindent
Our setup to determine the anomalous dimensions of four-fermion operators
in theories with adjoint fermions follows closely the approach
of~\cite{alpha},
where anomalous dimensions of four-fermion operators in QCD were studied
exploiting the Schr\"odinger Functional (SF) formalism.
The theory is regularised on a lattice of physical size $L^4$ with standard
SF Dirichlet boundary conditions, that allow to simulate at vanishing
fermion mass, and periodic boundary conditions for fermions in space
up to a phase $\theta$. Renormalisation conditions are imposed at a scale $\mu=1/L$.
To define them one first introduces SF correlation functions of the form
\vspace{-1mm}\begin{gather}
F_{i;{\rm A,B,C}}^\pm (x_0) = \frac{1}{L^3}\langle
\mathcal{O}'_{53}[\Gamma_{\rm C}] \mathcal{Q}_i^\pm \mathcal{O}_{21}[\Gamma_{\rm A}]\mathcal{O}_{45}[\Gamma_{\rm B}]
\rangle\,,
\end{gather}
where $\mathcal{Q}_i^\pm$ is the parity-odd part of $Q_i^\pm$; $\mathcal{O},\mathcal{O}'$ are interpolating fields on the time boundaries; and
five different choices for the Dirac matrices $\Gamma_{\rm A,B,C}$ are possible ---
see~\cite{alpha} for full details. Note that a fifth
``spectator'' flavour has been introduced.\footnote{In case the theory of
interest has less than five dynamical flavours, this setup can be interpreted
in terms of a valence sector used to the sole purpose of imposing renormalisation
conditions. Since all computations are carried out in the chiral limit, there
is no subtlety related to the matching of valence and sea fermions.}
The logarithmic divergence of $\mathcal{Q}_i^\pm$ can be isolated by dividing
out the divergencies coming from the boundaries, through ratios of the form
\vspace{-1mm}\begin{gather}
\label{eq:SFcor}
h_{i;{\rm A,B,C}}^\pm (x_0) = \frac{F_{i;{\rm A,B,C}}^\pm (x_0)}{f_1^\eta k_1^{3/2-\eta}}\,,
\end{gather}
where $f_1=-1/(2L^6)\langle\mathcal{O}'_{21}[\gamma_5]\mathcal{O}_{12}[\gamma_5]\rangle$ and $k_1=-1/(6L^6)\langle\mathcal{O}'_{21}[\gamma_k]\mathcal{O}_{12}[\gamma_k]\rangle$ are boundary-to-boundary correlators, and the value of $\eta\in[0,3/2]$ is fixed by the choice of $\Gamma_{\rm A,B,C}$.

Renormalisation conditions for the multiplicatively renormalisable operators
$\mathcal{Q}_1^\pm=Q_{\rm VA+AV}^\pm$ have the form
\vspace{-1mm}\begin{gather}
Z_1^\pm(g_0,a\mu)\,h^\pm_{1;{\rm A,B,C}}(L/2)=
\left.h^\pm_{1;{\rm A,B,C}}(L/2)\right|_{g_0=0}\,.
\end{gather}
For the other operators, that mix in doublets,
one has to impose matrix renormalisation conditions of the form
\vspace{-1mm}\begin{gather}
\left(\begin{array}{cc}
Z_{ii}^\pm(g_0,a\mu) & Z_{ij}^\pm(g_0,a\mu) \\
Z_{ji}^\pm(g_0,a\mu) & Z_{jj}^\pm(g_0,a\mu)
\end{array}\right)
\left(\begin{array}{c}
h_{i;{\rm A,B,C}}^\pm(L/2) \\
h_{j;{\rm A',B',C'}}^\pm(L/2)
\end{array}\right) =
\left(\begin{array}{c}
h_{i;{\rm A,B,C}}^\pm(L/2) \\
h_{j;{\rm A',B',C'}}^\pm(L/2)
\end{array}\right)_{g_0=0}\,.
\end{gather}
Different choices of boundary matrices result in different renormalisation
schemes.
In either case, sensible sets of renormalisation conditions require that
$Z^\pm$ is well-defined and equal to the identity at tree-level.

The equivalence of the formally different fierzings that appear in theories
with fundamental and adjoint fermions, alluded to above, can be easily
appreciated by working out the Wick contractions that enter the correlation
functions $h_{i;{\rm A,B,C}}^\pm$ in either case. Let us substitute
$\mathcal{Q}_i^\pm$ in Eq.~(\ref{eq:SFcor}) for either
$(\bar\psi_1\Gamma_1\psi_2)(\bar\psi_3\Gamma_2\psi_4)$,
$(\bar\psi_1\Gamma_1\psi_4)(\bar\psi_3\Gamma_2\psi_2)$, or
$(\bar\psi_1\Gamma_1C\bar\psi_3^T)(\psi_4^TC\Gamma_2\psi_2)$.
After integration over fermion fields, one has for the respective Wick
contractions expressions of the form
\vspace{-1mm}\begin{gather}
\begin{split}
&\langle
{\rm Tr}\{H_1(x)^\dagger\gamma_5\Gamma_1 H_2(x)\Gamma_{\rm A}\gamma_5\}
{\rm Tr}\{H_3'(x)^\dagger\gamma_5\Gamma_2 H_4(x)\Gamma_{\rm B}\gamma_5\mathcal{H}_5^\dagger\Gamma_{\rm C}\gamma_5\}
\rangle_{\rm\scriptscriptstyle G}\,,\\
-&\langle
{\rm Tr}\{H_1(x)^\dagger\gamma_5\Gamma_1 H_4(x)\Gamma_{\rm B}\gamma_5\mathcal{H}_5^\dagger\Gamma_{\rm C}\gamma_5H_3'(x)^\dagger\gamma_5\Gamma_2H_2(x)\Gamma_{\rm A}\gamma_5\}
\rangle_{\rm\scriptscriptstyle G}\,,\\
-&\langle
{\rm Tr}\{H_1(x)^\dagger\gamma_5\Gamma_1 H_{\tilde 4}(x)\Gamma_{\rm B}\gamma_5\mathcal{H}_{\tilde 5}^\dagger\Gamma_{\rm C}\gamma_5H_{\tilde 3}'(x)^\dagger\gamma_5\Gamma_2H_2(x)\Gamma_{\rm A}\gamma_5\}
\rangle_{\rm\scriptscriptstyle G}
\end{split}
\end{gather}
where $H,H'$ are bulk-to-boundary fermion propagators,
$\mathcal{H}$ is the boundary-to-boundary fermion propagator, and,
in the last expression, propagators bearing a tilde in the flavour index
are those for the fermion variables
$\tilde\psi(x) \equiv C\bar\psi(x)^T,\bar{\tilde\psi}(x) \equiv \psi(x)^T C$.
It is thus obvious that the structure of the correlation functions are identical
for either fermion representation.

Using standard finite size-scaling techniques, it is possible to compute
the renormalisation group running of the operators by computing renormalisation
constants on several lattices tuned so that the continuum limit can be
taken at fixed $\mu=1/L$ for several values of $\mu$, using a previous
determination of the SF running coupling $\overline{g}(\mu)$.
The basic objects to that purpose are the so-called step-scaling functions (ssf)
\vspace{-1mm}\begin{gather}
\Sigma^\pm(s;u,a/L) \equiv \left.
Z^\pm(g_0,a\mu/s)[Z^\pm(g_0,a\mu)]^{-1}
\right|_{u=\overline{g}^2(\mu)}\,,
\end{gather}
where $s>1$ is some fixed scaling parameter,
that can be defined either for multiplicatively renormalisable operators
or for operator doublets. The continuum limit of an ssf can be easily
written in terms of the continuum anomalous dimension matrix $\gamma^\pm$
and the $\beta$ function as
\vspace{-1mm}\begin{gather}
\sigma^\pm(s;u) \equiv \lim_{a\to 0}\Sigma^\pm(s;u,a/L)
= {\rm T}\exp\left\{
\int_{\overline{g}(\mu)}^{\overline{g}(\mu/s)}{\rm d} g\,
\frac{\gamma^\pm(g)}{\beta(g)}
\right\}\,,
\end{gather}
where the ${\rm T}\exp$ is meant to be ordered in $g$.
Close to an infrared fixed point, a good approximant for the
anomalous dimension is $\gamma^\pm(\overline{g}(\mu))\approx [\log\sigma^\pm(s;u)]/[\log s]$.

\vspace{-1mm}\section{Preliminary results in MWTC}

\noindent
Minimal Walking Technicolour --- i.e. an ${\rm SU}(2)$ gauge theory
with two flavours of fermions transforming in the adjoint colour representation ---
has been simulated at vanishing PCAC quark mass using the Hybrid MonteCarlo
algorithm for lattices of size
$L/a=8,10,12,16$ and $\beta=16.00,8.00,4.50,3.50,3.00,2.50,2.20,2.05$
for each lattice, using the Wilson plaquette action and the unimproved
Wilson fermion action. SF correlation functions for four-fermion operators,
as well as quark bilinears (which allows to study the running techniquark mass,
providing a cross-check with and extending the study of~\cite{MWTC1}),
have been computed at vanishing background field and a value of $\theta=0.5$
for the abelian spatial twist. Approximately 20000 thermalised trajectories
are available at each simulation point, with acceptance rates ranging
between $83\%$ and $96\%$.

\begin{figure}[!t]
\vspace*{-5mm}
\hspace*{0mm}\includegraphics[scale=0.42]{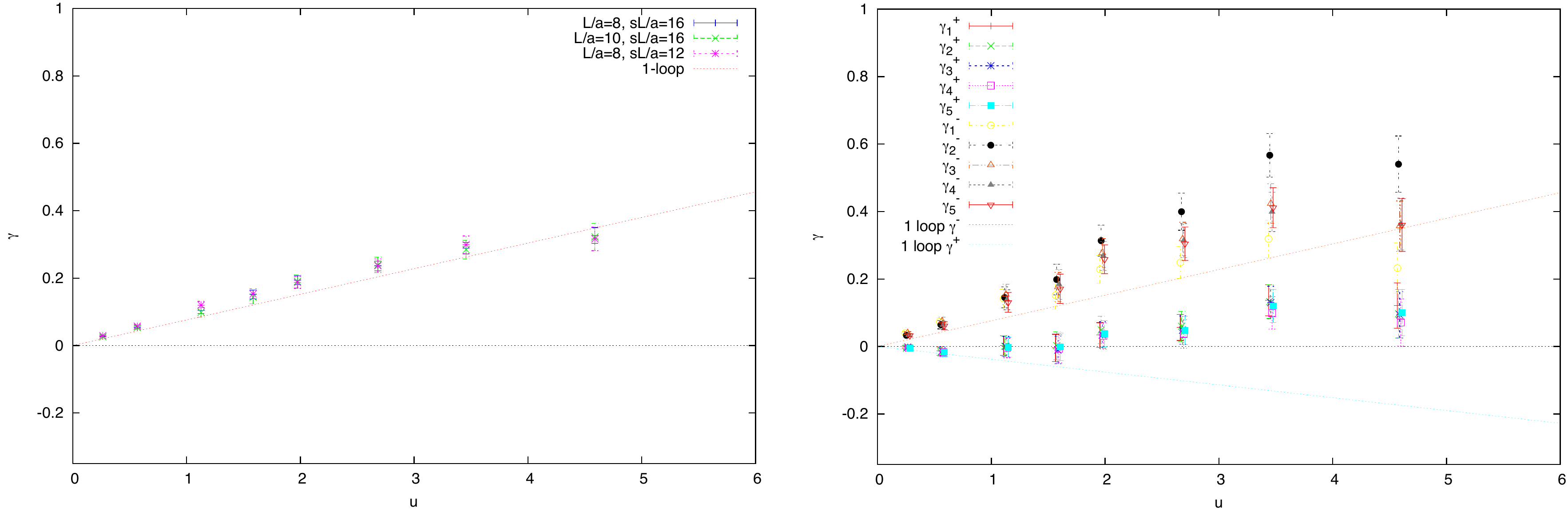}
\\\vspace{-9mm}
\caption{Left: anomalous mass dimension for various scaling steps. Right: anomalous dimensions of $Q_{\rm VA+AV}^\pm$ from $L/a=8~\to~16$.}
\label{fig1}
\end{figure}

\begin{figure}[!t]
\vspace*{-0mm}
\hspace*{0mm}\includegraphics[scale=0.42]{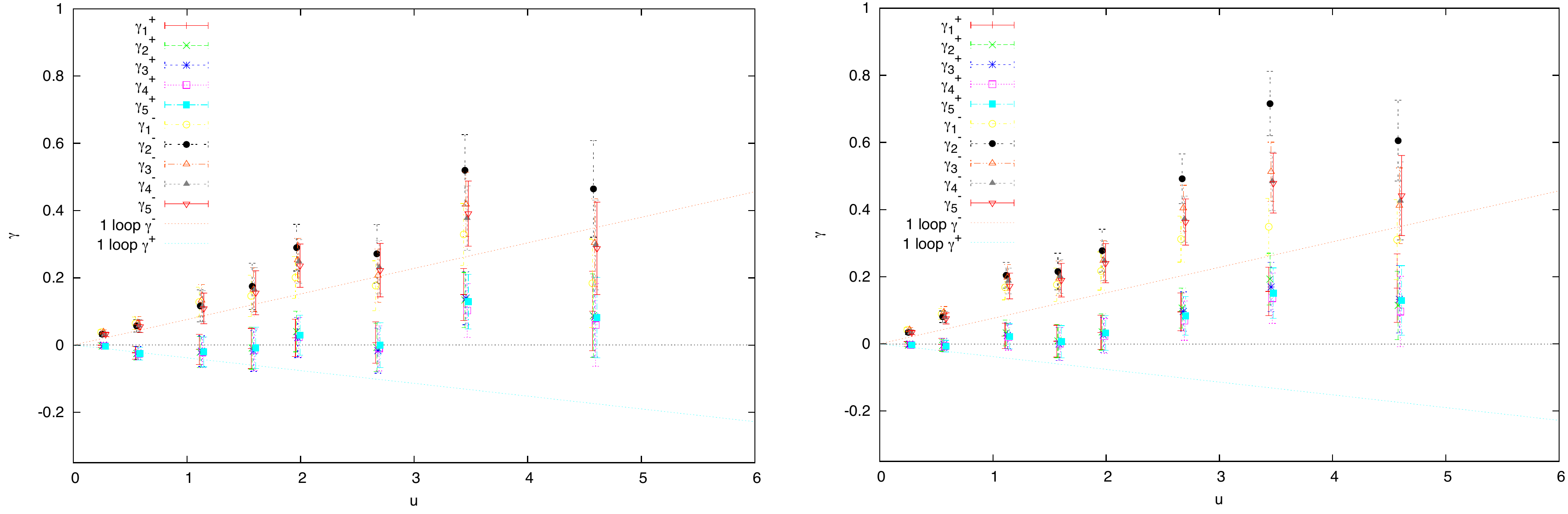}
\\\vspace{-9mm}
\caption{Left: anomalous dimensions of $Q_{\rm VA+AV}^\pm$ from $L/a=10~\to~16$. Right: idem from $L/a=8~\to~12$.}
\label{fig2}
\end{figure}

The left panel of Fig.~\ref{fig1} shows (minus) the quantity $\log\sigma(s;u)/\log s$ for the pseudoscalar density (i.e. the mass anomalous dimension)
for various step scalings. The right panel
shows the same quantity for the multiplicatively renormalisable operators
$Q^\pm_{\rm VA+AV}$ and fixed step scaling from $L/a=8$ to $L/a=16$.
In the latter case the five available renormalisation schemes are shown.
The values of $\beta$ have been converted to values of
$u=\overline{g}^2$ based on the running coupling data at $L=8$, and
the one-loop predictions are also shown on the plots for comparison.
While the results for the four-fermion anomalous dimensions are much
less precise, owing to the use of four-point correlation functions and
the intrinsically noisier four-fermion insertion, they display a
remarkable independence of the scheme choice, as expected for a theory
that displays conformal behaviour and quite contrary to the findings
in QCD~\cite{alpha}.
Within the large errors, around the values of the coupling
for which the appearance of the fixed point is suspected one has
$\gamma^+ \sim 0,~\gamma^- \sim 0.4$. The mild dependence on the
choice of the step scaling is illustrated in Fig.~\ref{fig2},
which confirms the rough consistence with the expectations coming
from conformal dynamics. Preliminary results for the anomalous dimension
matrices involving the other four operators, which include the phenomenologically
interesting case $SP + PS$, are also consistent with this picture.

\vspace{-1mm}\section{Outlook}

\noindent
Studying the anomalous dimensions of four-fermion operators is
an important step in the understanding the phenomenological implications
of strongly coupled models for EWSB. We have reviewed the general
framework to determine anomalous dimensions non-perturbatively,
and extended it to theories with adjoint fermions.

Our MWTC results are quite preliminary, not least because of the uncontrolled
systematic uncertainties stemming from not having a continuum limit,
a significant statistical error, and our uncertain knowledge about the
location of the (possible) fixed point. Yet, our results show a non-trivial
consistency with expectations from conformal dynamics, and provide ballpark
values for the anomalous dimensions in the would-be conformal region.
After our analysis is complete, including results for the full basis of
operators, a next obvious step is to exploit the ample margin for better
precision offered by increasing our statistics, simulating larger lattices
to allow for a better study of the approach to the continuum limit,
or implementing $\mathcal{O}(a)$ improvement. Extension to other models
with adjoint or fundamental fermions is straightforward.


\begin{thebibliography}{99}

\bibitem{TC}
  S.~Weinberg,
  Phys.\ Rev.\ D {\bf 19} (1979) 1277;
  L.~Susskind,
  Phys.\ Rev.\ D {\bf 20} (1979) 2619.

\bibitem{Holdom:1981rm}
  B.~Holdom,
  Phys.\ Rev.\ D {\bf 24} (1981) 1441.

\bibitem{WTC}
  K.~Yamawaki, M.~Bando and K.-I.~Matumoto,
  Phys.\ Rev.\ Lett.\  {\bf 56} (1986) 1335;
  T.~Akiba and T.~Yanagida,
  Phys.\ Lett.\ B {\bf 169} (1986) 432;
  T.W.~Appelquist, D.~Karabali and L.C.R.~Wijewardhana,
  Phys.\ Rev.\ Lett.\  {\bf 57} (1986) 957.

\bibitem{kuti}
  J.~Kuti, these proceedings.

\bibitem{Rattazzi:2008pe}
  R.~Rattazzi, V.~S.~Rychkov, E.~Tonni and A.~Vichi,
  JHEP {\bf 0812} (2008) 031.

\bibitem{MWTC1}
 F.~Bursa et al.,
 Phys.\ Rev.\  D {\bf 81}, 014505 (2010).

\bibitem{MWTC2}
 L.~Del Debbio et al.,
 Phys.\ Rev.\  D {\bf 82}, 014509 (2010);
 L.~Del Debbio et al.,
 Phys.\ Rev.\  D {\bf 82}, 014510 (2010);
 T.~DeGrand, Y.~Shamir and B.~Svetitsky,
 Phys.\ Rev.\  D {\bf 83}, 074507 (2011);
 S.~Catterall, L.~Del Debbio, J.~Giedt and L.~Keegan,
 Phys.\ Rev.\  D {\bf 85}, 094501 (2012).

\bibitem{Donini:1999sf}
  A.~Donini et al.,
  Eur.\ Phys.\ J.\ C {\bf 10} (1999) 121

\bibitem{Testa:1998ez}
  M.~Testa,
  JHEP {\bf 9804} (1998) 002.

\bibitem{alpha}
  M.~Guagnelli {\it et al.}  [ALPHA Collaboration],
  JHEP {\bf 0603} (2006) 088;
  F.~Palombi, C.~Pena and S.~Sint,
  JHEP {\bf 0603} (2006) 089;
  P.~Dimopoulos {\it et al.}  [ALPHA Collaboration],
  JHEP {\bf 0805} (2008) 065.

  
\end{thebibliography}
\end{document}